\documentclass[aps,prb,twocolumn,superscriptaddress,floatfix,showpacs,10pt,letterpaper]{revtex4-1}

\pdfoutput=1

\usepackage{graphicx}
\usepackage{amsmath}
\usepackage{amstext}
\usepackage{amssymb}
\usepackage{amsfonts}
\usepackage{amsbsy} 
\usepackage{dcolumn}
\usepackage{bm}
\usepackage{multirow}

\usepackage{color}
\usepackage[colorlinks,citecolor=blue]{hyperref}

\usepackage[bbgreekl]{mathbbol}
\usepackage{amscd}

\newcommand{\<}{\langle} 
\renewcommand{\>}{\rangle}

\newcommand{\cT}{ \mathcal{T} }

\newcommand{\bpm}{\begin{pmatrix}}
\newcommand{\epm}{\end{pmatrix}}
\newcommand{\bmm}{\begin{matrix}}
\newcommand{\emm}{\end{matrix}}
\newcommand{\be}{\begin{equation}}
\newcommand{\ee}{\end{equation}}

\usepackage{graphicx}% Include figure files
\usepackage{dcolumn}% Align table columns on decimal point
\usepackage{bm}% bold math
\usepackage{epstopdf}
\usepackage{setspace}

\newcommand{\beq}{\begin{equation}}
\newcommand{\eeq}{\end{equation}}
\newcommand{\ba}{\begin{array}{ccc}}
\newcommand{\ea}{\end{array}}
\newcommand{\nn}{\nonumber}
 \renewcommand{\d}{\partial}
\def\bea{\begin{eqnarray}}
\def\eea{\end{eqnarray}}

\def\<{\langle}
\def\>{\rangle}

\usepackage{amsmath}
\usepackage{amssymb}

\begin{document}

\title{Time reversal invariant gapped boundaries of the double semion state}

\author{Fiona Burnell}
\affiliation{Department of Physics and Astronomy, University of Minnesota, Minneapolis, MN 55455, USA}
\author{Xie Chen}
\affiliation{Department of Physics and Institute for Quantum Information and Matter, California Institute of Technology, Pasadena, CA 91125, USA}
\affiliation{Department of Physics, University of California, Berkeley, CA, 94720, USA}
\author{Alexei Kitaev}
\affiliation{Department of Physics and Institute for Quantum Information and Matter, California Institute of Technology, Pasadena, CA 91125, USA}
\author{Max Metlitski}
\affiliation{Kavli Institute for Theoretical Physics, UC Santa Barbara, CA 93106, USA}
\author{Ashvin Vishwanath}
\affiliation{Department of Physics, University of California, Berkeley, CA, 94720, USA}
\affiliation{Materials Science Division, Lawrence Berkeley National Laboratories, Berkeley, CA 94720, USA}

\begin{abstract}
The boundary of a fractionalized topological phase can be gapped by condensing a proper set of bosonic quasiparticles. Interestingly, in the presence of a global symmetry, such a boundary can have different symmetry transformation properties. Here we present an explicit example of this kind, in the double semion state with time reversal symmetry. We find two distinct cases where the semionic excitations on the boundary can transform either as time reversal singlets or as time reversal (Kramers) doublets, depending on the coherent phase factor of the Bose condensate. The existence of these two possibilities are demonstrated using both field theory argument and exactly solvable lattice models. Furthermore, we study the domain walls between these two types of gapped boundaries and find that the application of time reversal symmetry tunnels a semion between them. 
\end{abstract} 

\maketitle

%{\small \setcounter{tocdepth}{1} \tableofcontents }

%%%%%%%%%%%%%%%%%%%%%%%%%%%%%%%%%%%%%%%%%%%%%%%%%%%%%%%%%%%%%%%%%%%%%%%%%%%%%%%

\section{Introduction}

The interplay of topology and symmetry can lead to interesting phenomena in quantum many-body systems. In particular, in the presence of global symmetries, one topological phases can divide into several different phases with the fractional excitations in the system transforming under symmetry in different ways. Much recent effort has been devoted to the classification of such `Symmetry Enriched Topological' (SET) phases by identifying possible ways for the symmetry to act on the fractional excitations\cite{Wen2002,Levin2012,Essin2013,Mesaros2013,Hung2013,Lu2013,Xu2013,Chen2014,Gu2014,Barkeshli2014,Fidkowski}. One possibility is for the fractional excitations to carry fractional quantum numbers of the global symmetry. For example, in an electronic system composed of charge $e$ electrons, the fractional excitations in the $\nu=1/3$ fractional quantum Hall state can carry charge $e/3$. 
A systematic counting exists for this class of SET phases\cite{Essin2013} (although it is not completely clear which of these phases exist in pure two dimension and which exist as the surface of a three dimensional system). 

Another possibility in SET phases is for the symmetry to map one type of fractional excitation to another. %A similar question exists as to how many SET phases there are in this class. 
For example, the double semion topological order, which exists in for example a double layer fractional quantum Hall system with $\nu=\pm 1/2$, is time reversal invariant. Time reversal symmetry maps between the semion (with topological spin $i$) and the anti-semion (with topological spin $-i$) while keeping their combination -- a bosonic quasiparticle -- invariant. 

\begin{figure}[htbp]
\centering
\includegraphics[width=2in]{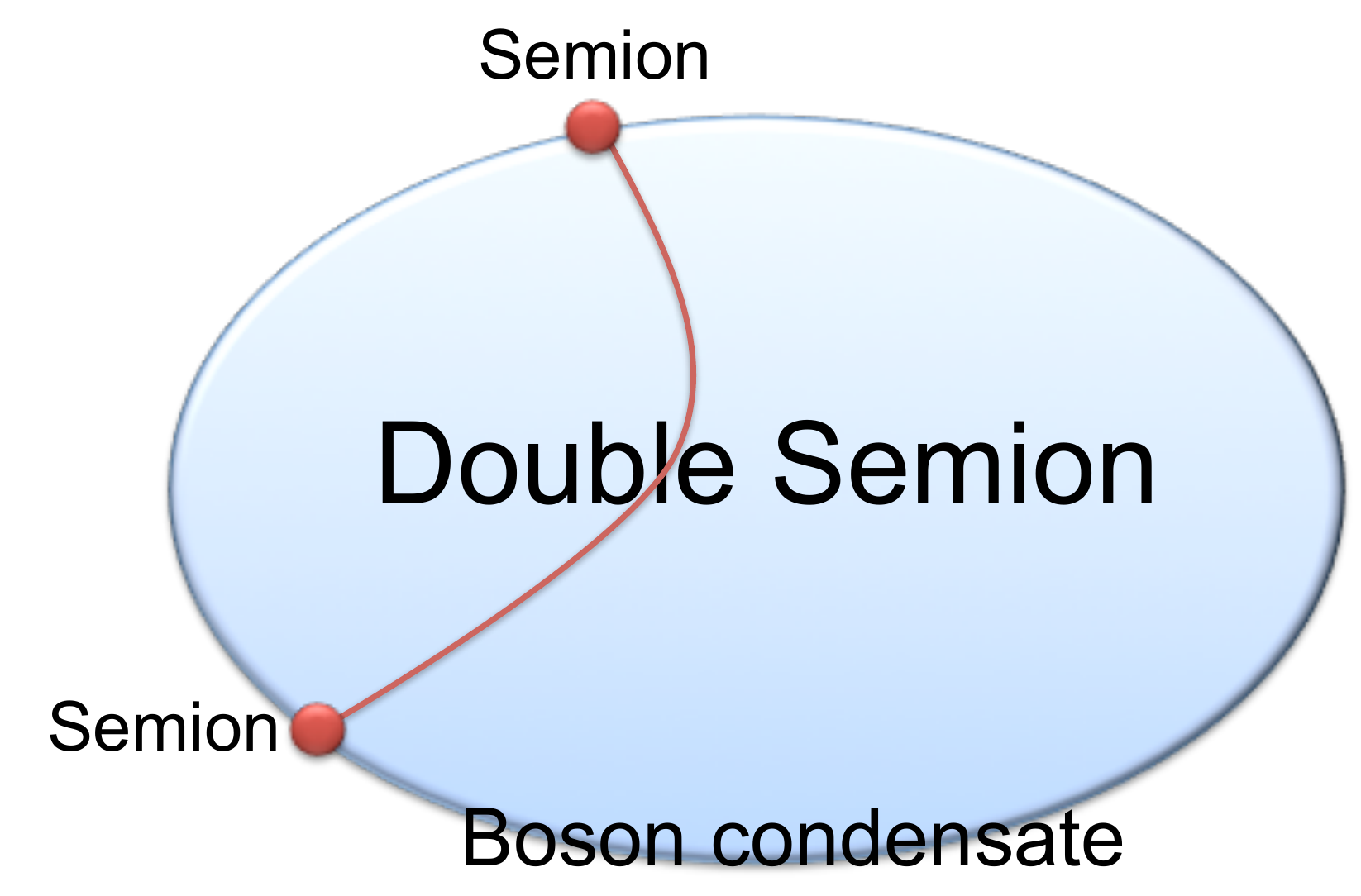}
\caption{A double semion state whose boundary is gapped by condensing the bosonic quasiparticle.}
\label{DS}
\end{figure}

Even though the semion / anti-semion are not individually invariant under time reversal symmetry, one might wonder if we can still make sense of the `quantum number' or local symmetry transformation on each of them. This question becomes more concrete when we consider the boundary of the topological state. The boundary of the double semion state can be gapped, as shown in Fig.\ref{DS}, by condensing the bosonic quasiparticle, which is the combination of the semion and the anti-semion. Because both the semion and the anti-semion obtain nontrivial phase factors ($-1$) when braiding around the boson, they become confined in the condensate. Moreover, because the bosonic quasiparticle is condensed, i.e. they can appear and disappear freely, a semion becomes indistinguishable from an anti-semion. Now it becomes reasonable to ask about the local time reversal symmetry transformation of the semion (anti-semion) and there could be two options: $\cT^2=1$ and $\cT^2=-1$. Such a distinction leads to measurable physical effect in the system. Imagine creating a pair of semions in the bulk, separating them from each other and bringing them to the gapped boundary. When the semion transforms as $\cT^2=-1$, a local Kramer degeneracy appears at the location of each semion while with $\cT^2=1$ semion no local degeneracy is expected. 

So what does this `time reversal quantum number' of the semion imply? Does it label different bulk SET phases or does it correspond to different boundary conditions for the same bulk phase? In this paper we show that the latter is true. In particular, we use both field theory arguments and exactly solvable models to show that there are two types of Bose condensates in the double semion state, with different coherent phase factors of the condensed boson. In one of them the semion transforms as $\cT^2=1$ while in the other it transforms as $\cT^2=-1$.

The paper is organized as follows: In section \ref{FT}, we present a simple field theory argument for the result, which is supported by exactly solvable models constructed in section \ref{ESM}. In section \ref{DWall}, we consider the situation where different segments of the boundary of the double semion state is gapped in the two different ways and ask what happens at the domain wall between the segments. We find that the domain wall carries extra degeneracy protected by time reversal symmetry and the symmetry action tunnels a semion between pairs of domain walls. In section \ref{SPT}, we discuss how this is all related to the symmetry protected topological phase with $Z_2$ and time reversal symmetry, which becomes the double semion state under study by gauging the $Z_2$ symmetry. Finally, we conclude in section \ref{Conclusion} and compare this example with similar models studied previously.

%%%%%%%%%%%%%%%%%%%%%%%%%%%%%%%%%%%%%%%%%%%%%%%%%%%%%%%%%%%%%%%%%%%%%%%%%%%%%%%

\section{Field theory analysis}
\label{FT}

The double semion state contains an abelian topological order with three types of fractional excitations: the semion $s$, the anti-semion $s'$ and their combination -- the boson $b=ss'$. The topological spins of the three are $i$, $-i$ and $1$ respectively and the mutual statistics between $s$ and $s'$ is trivial. In field theory language, the double semion topological order can be described as a $U(1)\times U(1)$ Chern-Simons theory: 
\be
L = \frac{2}{4\pi} \epsilon^{\lambda\mu\nu}a^1_{\lambda}\partial_{\mu}a^1_{\nu} - \frac{2}{4\pi} \epsilon^{\lambda\mu\nu}a^2_{\lambda}\partial_{\mu}a^2_{\nu}
\ee
whose edge state can described as
\be
L_e = \frac{2}{4\pi} \partial_x \phi_1 \partial_t \phi_1 - \frac{2}{4\pi} \partial_x \phi_2 \partial_t \phi_2
\label{Le}
\ee
where only the topological term in the Lagrangian is shown.

Time reversal symmetry action on the edge fields $\phi_1$ and $\phi_2$ can be written in two ways:
\be
\cT_1: \phi_1 \rightarrow \phi_2, \phi_2 \rightarrow \phi_1
\ee
Or equivalently,
\be
\cT_2: \phi_1 \rightarrow \phi_2, \phi_2 \rightarrow \phi_1+\pi
\ee
$\cT_1$ and $\cT_2$ differ by a gauge transformation\cite{Wen1995}
\be
g: \phi_1 \rightarrow \phi_1+\pi, \phi_2 \rightarrow \phi_2
\ee
$g$ acts as $g=(-1)^{N_s}$ where $N_s$ is the number of semions on the edge and the action is trivial on all local operators of the form $e^{i2n\phi_1+i2m\phi_2}$ with integer $n$ and $m$.

It appears that a semion, generated by $e^{i\phi_1}$ (or the anti-semion generated by $e^{i\phi_2}$) transforms as $\cT^2=1$ under $\cT_1$ and as $\cT^2=-1$ under $\cT_2$. So which time reversal transformation should we use? That depends on the boundary condition we choose for the edge theory. In particular, the edge state described by Eq.\ref{Le} can be gapped out by adding a Higgs term
\be
\Delta L = -\lambda \cos(2\phi_1-2\phi_2+\alpha)
\ee
with $\lambda>0$. When $\lambda$ is large enough, the bosonic quasiparticle, generated by $e^{i(\phi_1-\phi_2)}$, is condensed on the edge. However, there are two types of condensates which preserve time reversal symmetry, one with $\alpha=0$ and the other with $\alpha=\pi$. 

When $\alpha=0$, the term $\Delta L$ has two classical minima:
\be
\phi_1-\phi_2 = 0 , \ \ \phi_1-\phi_2 = \pi
\ee
The two minima are related by $g$ and hence are physically identical (no local observable distinguishes them). Each minimum is invariant under $\cT_1$. Therefore this is the form of time reversal transformation that we should consider and the semion $e^{i\phi_1}$ (or the anti-semion $e^{i\phi_2}$) transforms as time reversal singlets $\cT^2=1$ on the boundary.

When $\alpha=\pi$, the term $\Delta L$ has two classical minima as well:
\be
\phi_1-\phi_2 = \pi/2, \ \ \phi_1-\phi_2 = -\pi/2
\ee
The two minima are related by $g$ again but neither of them is invariant under $\cT_1$. Instead they are preserved by $\cT_2$. Therefore, $\cT_2$ is the manifest time reversal operation on the edge when boson is condensed with $\alpha=\pi$ and the semion $e^{i\phi_1}$ (or the anti-semion $e^{i\phi_2}$) transforms as time reversal doublets $\cT^2=-1$ on the boundary.

Let us look more carefully at how the time reversal transformation of the semion / anti-semion depends on the coherent phase factor of the Bose condensate. Note that in the condensates, the process of creating or annihilating a boson pair is associated with a phase factor of
\be
<e^{i(\phi_1(x)-\phi_2(x))}e^{i(\phi_1(x')-\phi_2(x'))}> = e^{i\alpha} = \pm 1
\ee
while the process of boson hopping always comes with a phase factor of $1$
\be
<e^{i(\phi_1(x)-\phi_2(x))}e^{-i(\phi_1(x')-\phi_2(x'))}> = 1
\ee 
Therefore, the total wave function of the condensate reads
\be
|\psi_{bc}\rangle = \sum_{N} (e^{i\alpha})^N \sum_{x_1,...,x_{2N}} |x_1,...,x_{2N}\rangle
\ee
where the inner sum is over all possible position configurations of $2N$ bosons and the outer sum is over all integer $N$. Note that there are always an even number of bosons because it is a quasiparticle (self) boson and can only be created in pairs. 

The relation between the condensed phase and the time reversal transformation of the semions can be understood as follows: when applying time reversal to a semion, it is mapped to an anti-semion and hence a boson is created; when applying time reversal again and mapping the anti-semion back to a semion, another boson is created. That is, the process of applying $\cT^2$ to a semion is accompanied by the creation of a boson pair and hence
\be
\cT^2 = e^{i\alpha}
\ee
on each semion.

%%%%%%%%%%%%%%%%%%%%%%%%%%%%%%%%%%%%%%%%%%%%%%%%%%%%%%%%%%%%%%%%%%%%%%%%%%%%%%%

\section{Exactly solvable model construction}
\label{ESM}

Using the exactly solvable model of double semion introduced by Ref.\onlinecite{Freedman2004,Levin2005}, we can construct the two types of condensates corresponding to $\alpha=0$ and $\alpha=\pi$ and show explicitly that the semion excitations transform respectively  as $\cT^2=1$ and $\cT^2=-1$ under time reversal.

\subsection{$\alpha=0$ condensate and $\cT^2=1$ semion}

Consider the double semion model on Honeycomb lattice, with one spin $1/2$ living on each link. To simplify notation, we write spin operators $\sigma_x$, $\sigma_y$, $\sigma_z$ as $X$, $Y$, $Z$. 
The Hamiltonian in the topological bulk contains vertex terms $A_v$ and plaquette terms $B_p$\cite{Levin2005}.
\be
H_T = -\sum_v A_v + \sum_p B_p
\ee
where 
\be
\begin{array}{l}
A_v = \prod_{i\in v} Z_i \\
B_p = \prod_{i \in p} X_i \prod_{j \in p} (-)^{n_j(1-n_{j+1})}
\end{array}
\ee 
$i\in v$ labels links attached to a vertex $v$ and $j\in p$ labels links around a plaquette $p$.
$n_j=0$ if $Z_j=1$ and $n_j=1$ if $Z_j=-1$. 

Time reversal $\cT$ acts as complex conjugation in the $Z$ basis. Then the Hamiltonian is time reversal invariant and so is the ground state. Note that the form of the plaquette term $B_p$ used here is slightly different from that used in Ref.\onlinecite{Levin2005}, but they are equivalent when all vertex terms $A_v$ are satisfied. The form used here is simpler for our purpose because it is explicitly time reversal invariant.

It is easy to check that 1. all $A_v$'s commute, all $A_v$'s commute with all $B_p$'s and $B_p$'s commute with each other when the $A_v$ constrains are satisfied 2. $(B_p)^2=1$.

\begin{figure}[htbp]
\centering
\includegraphics[width=2.5in]{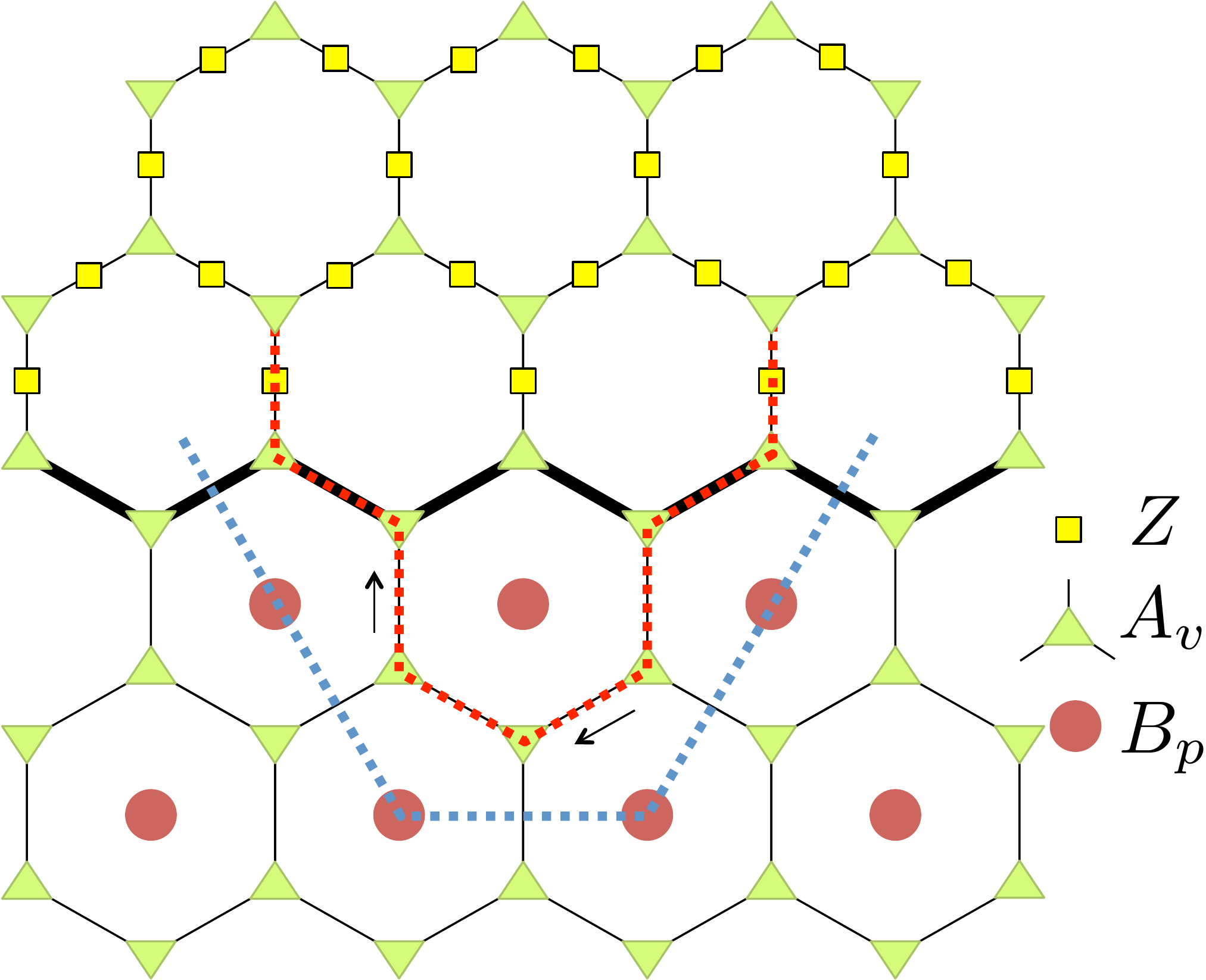}
\caption{Boundary (thick black line) of the double semion state (lower half plane) gapped by $\alpha=0$ Bose condensate (upper half plane) with the semion excitation transforming as $\cT^2=1$ under time reversal.}
\label{C0}
\end{figure}

Now let's condense the bosonic quasiparticle in the double semion state and gap out the boundary. In the simplest form, the boson condensation can be achieved by enforcing a $Z_k$ term on the link variables. That is
\be
H_{C0} = -\sum_k Z_k - \sum_v A_v
\ee
Obviously the second term is redundant. We include it here just for comparison with later cases. 
In the double semion state, the $Z_k$ term creates / annihilates boson pairs and also hops bosons around. Therefore, when the $H_{C0}$ term dominates over the $H_T$ term, the boson is condensed. Here we consider the exactly solvable situation as shown in Fig.\ref{C0} where the Hamiltonian in the region below the thick black boundary line is $H_T$ and that above the boundary line is $H_{C0}$. All Hamiltonian terms commute with each other.

From simple counting, we see that the boundary between the condensate and the topological region is totally gapped. Moreover, this condensate does not break time reversal symmetry. A boson-boson pair is created by string operator
\be
W_{1}^b = \prod_{k \in L'} Z_k
\ee
with $L'$ being a string in the dual lattice. $W_{1}^b$ obviously has eigenvalue $1$ everywhere in the condensate. Boson hopping is also generated with this term and also comes with a phase factor of $1$. Therefore, the condensate generated with $H_{C0}$ is the $\alpha=0$ condensate discussed in the previous section.

Now let's see what happens when we create a semion on the boundary. A semion-semion pair is created, as shown in Fig.\ref{C0} along the red dotted line $L$. We have chosen a particular direction for this string operator. %In the topological region, 
The string operator acts as\cite{Levin2005}
\be
W^s = \prod_{k \in L} X_k \prod_{v_l} \alpha_l \prod_{v_r} \alpha_r
\ee
where $\alpha_l$ and $\alpha_r$ are phase factors in the $Z$ basis at vertices where the string turns left ($v_l$) and right ($v_r$) respectively. $\alpha_l=\pm 1$ is always real. $\alpha_r = (i)^{n_r}$ acts on the leg to the left side of $L$ at this vertex. Obviously in the condensate, the string operator costs linear energy and the semion is confined.

Under time reversal symmetry, this string operator changes. In particular, the $\alpha_r$ phase factors change to $\alpha^{-1}_r$ and the difference is
\be
W_{1}^b = \prod_{k \in L'} Z_k
\ee
where $L'$ is the string in the dual lattice to the left side of $L$, as shown in Eq.\ref{C0} with the dotted blue line. This is nothing but the string operator for creating a pair of bosons on the boundary, which has eigenvalue $1$. Therefore, the state with a pair of semions on the boundary is invariant under time reversal and each semion transforms as a time reversal singlet with $\cT^2=1$.

\subsection{$\alpha=\pi$ condensate and $\cT^2=-1$ semion}

\begin{figure}[htbp]
\centering
\includegraphics[width=2.5in]{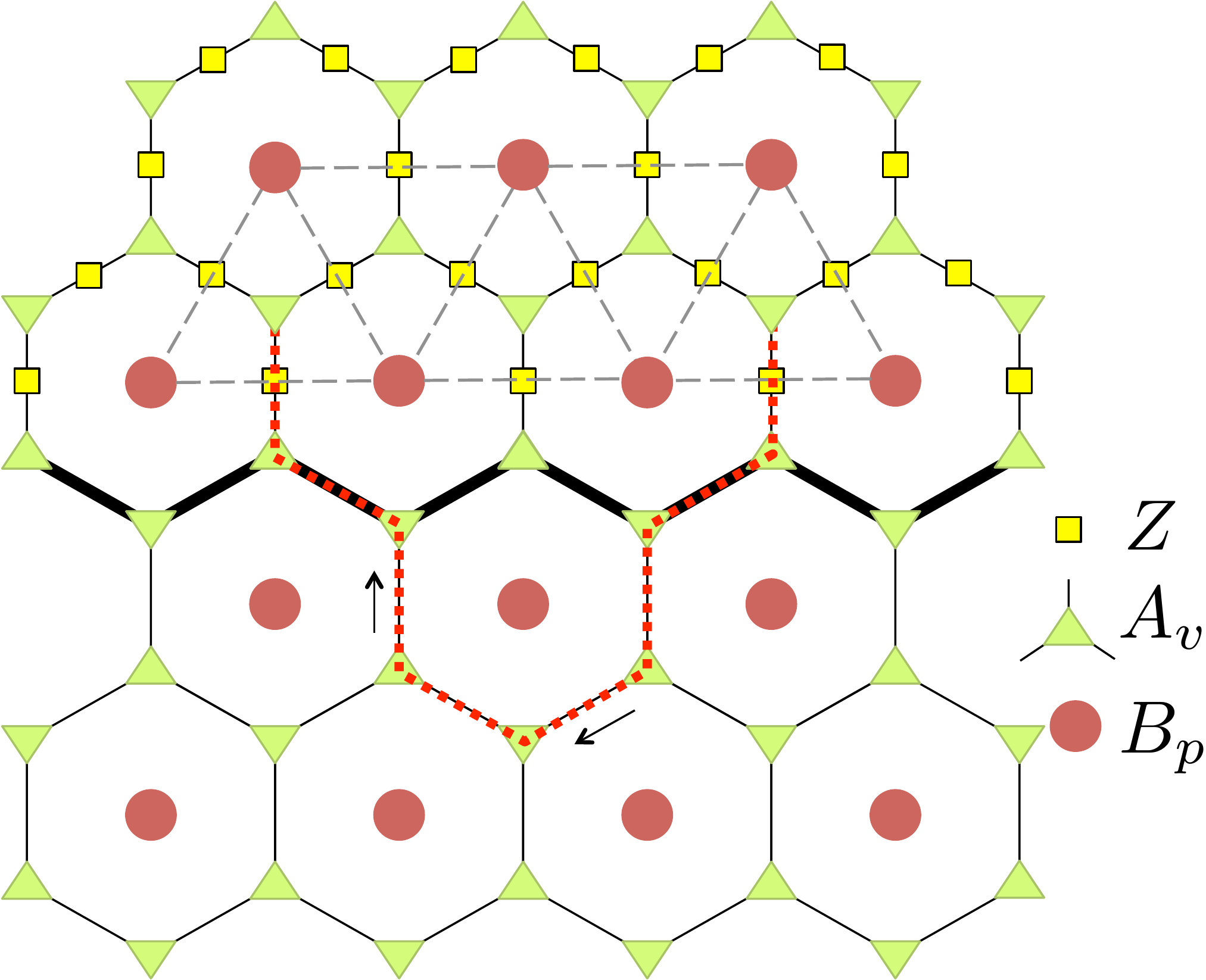}
\caption{Boundary (thick black line) of the double semion state (lower half plane) gapped by $\alpha=\pi$ Bose condensate (upper half plane) with the semion excitation transforming as $\cT^2=-1$ under time reversal. The Hamiltonian terms in the condensate $B_{p_a}Z_{ab}B_{p_b}$ are represented as two red dots ($B_{p_a}$ and $B_{p_b}$) connected by a grey dashed line via a yellow triangle ($Z_{ab}$).}
\label{Cpi}
\end{figure}

Now let's define a second type of condensate and see how the semion on the boundary can transform as a time reversal doublet. The Hamiltonian realizing the second type of condensate is
\be
H_{C\pi} = - \sum_{<ab>} B_{p_a}Z_{ab}B_{p_b} - \sum_v A_v
\label{HII}
\ee
where $B_{p_a}$ and $B_{p_b}$ are neighboring plaquette operators sharing a link $ab$ and the sum is over all such pairs. In the subspace where all vertex constraints $A_v$ are satisfied, it is easy to see that all terms in $H_{C\pi}$ commute with each other ($Z_{ab}$ anti-commutes with $B_{p_a}$ or $B_{p_b}$). Moreover
\be
(B_{p_a}Z_{ab}B_{p_b})^2 = 1
\ee
The relative signs in Eq.\ref{HII} are fixed by the requirement that the Hamiltonian is not frustrated. With our choice of signs it is not, since
\be
1=A_v =(B_{p_1}Z_{12}B_{p_2})(B_{p_2}Z_{23}B_{p_3})(B_{p_3}Z_{31}B_{p_1})
\ee
when both $B_{p_a}Z_{ab}B_{p_b}=1$ and $A_v=1$.

The $B_{p_i}Z_{ij}B_{p_j}$ terms create boson pairs or hop bosons around in this condensate. %The action of this term is such that it adds an extra phase factor when a boson quasiparticle (as a $B_P=-1$ excitation) already exists and is moved around by this term compared to the situation when boson quasiparticles do not exist $B_p=1$ and boson pairs are created / annihilated by this term. 
When this term acts on a pair of plaquettes with either 2 or 0 bosons ($B_P1 B_P2 =1$), there is an extra sign relative to its action on a pair of plaquettes with only one boson ($B_P1 B_P2 =-1$).  Counting the phase factors at both ends, pair creation / annihilation is accompanied by a phase factor of $-1$ while boson hopping has a phase factor of $+1$. Note that this is true even when we put the small pieces of boson string operators together and make longer strings. 
\be
\begin{array}{l}
(B_{p_1}Z_{12}B_{p_2})(B_{p_2}Z_{23}B_{p_3})...(B_{p_{m-1}}Z_{(m-1)m}B_{p_m}) \\
= B_{p_1}Z_{12}...Z_{(m-1)m}B_{p_m}
\end{array}
\ee
Hence in the ground state we have
\be
B_{p_1}Z_{12}...Z_{(m-1)m}B_{p_m} = 1
\ee
Therefore, the condensate generated with $H_{C\pi}$ corresponds to the $\alpha=\pi$ condensate discussed in section \ref{FT}.
%Therefore, the wave function of the Bose condensate can be written as\
%\be
%|\psi\rangle = \sum_{N} \sum_{x_1,...,x_{2N}} e^{i\pi N} |b_1(x_1),...,b_{2N}(x_{2N})\rangle
%\ee
Even though an extra phase factor is present in the wave function of this condensate, the condensate is still time reversal invariant. % Note that the total number of bosons in the condensate is always even because the boson is a fractional excitation and can only appear in pairs.

Now let's create a semion pair on the boundary along the red dotted line in Fig.\ref{Cpi} using the same string operator $W^s$ as in the previous section. Similar to the previous case, the string operator $W^s$ violates the vertex $A_v$ terms at its end points, but this has extra consequences in this new type of condensate. In particular, when $A_v$ is violated, the plaquette operators $B_p$ around this vertex no longer commute with each other. To restore exact solvability, certain terms need to be removed, introducing local degeneracies into the low energy Hilbert space. As we will show below, this local degeneracy indicates the presence of a local Kramers pair with $\cT^2=-1$ under time reversal.

\begin{figure}[htbp]
\centering
\includegraphics[width=2in]{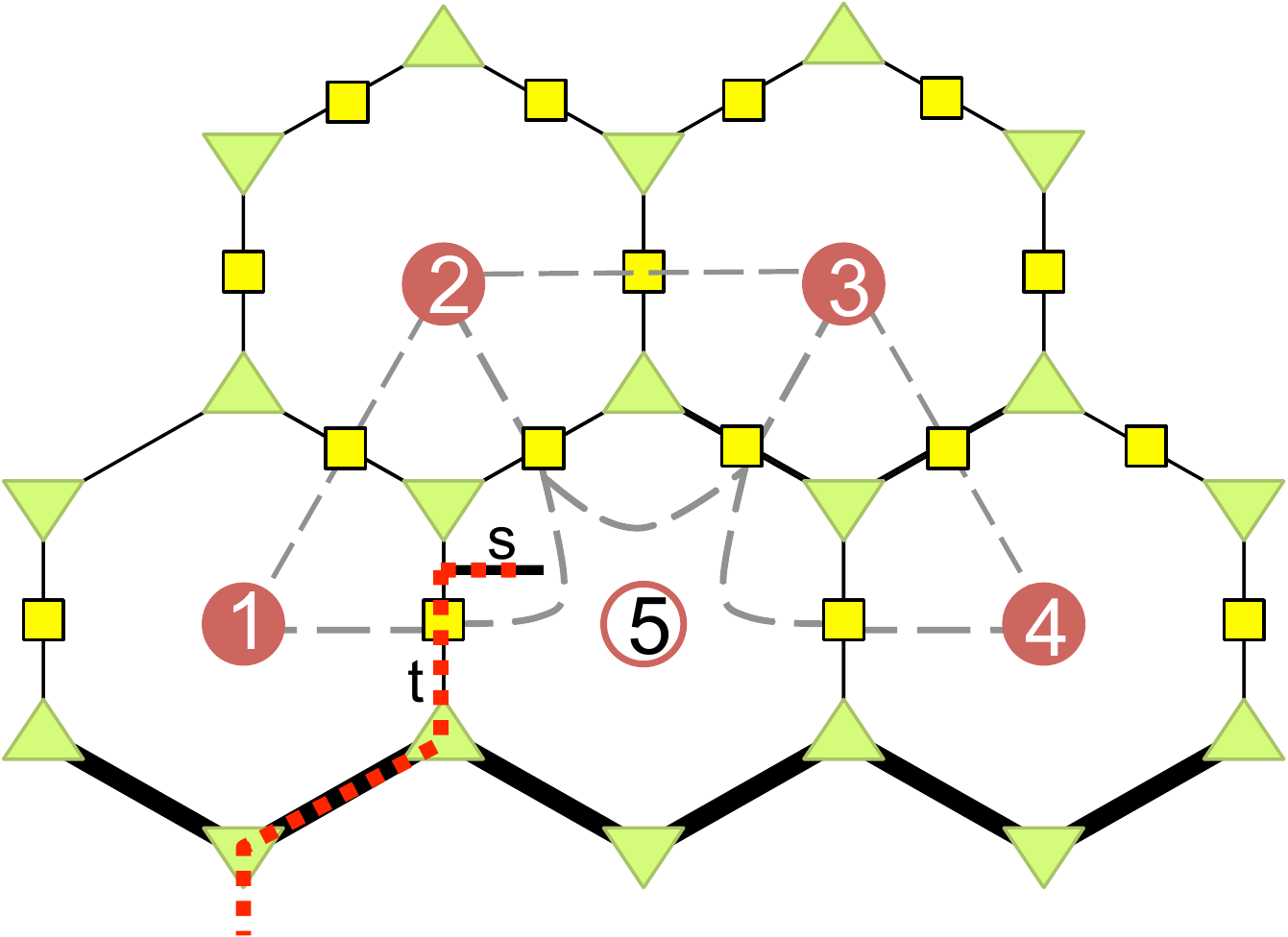}
\caption{Semion excitation on the boundary with $\alpha=\pi$ condensate transforms as a Kramer doublet with $\cT^2=-1$.}
\label{doublet}
\end{figure}

Let us zoom in on the semion as shown in Fig.\ref{doublet}. To expose the two fold Kramer degeneracy related to the semion, we redefine the Hamiltonian locally as follows: imagine breaking the link between plaquette $1$ and $5$ into two parts and adding a link $s$ sticking into plaquette $5$ as shown in Fig.\ref{doublet}. The state of the link $s$ is initially set to be $n_s=0$. We can choose the semion string operator as going into the condensate and ending on link $s$. The state of the link $s$ is hence flipped to $n_s=1$. By doing so, we have moved the vertex violation to the end of link $s$ which does not affect plaquette $1$ and $2$ but only plaquette $5$. 

Due to the existence of link $s$, $B_{p_5}$ needs to be redefined. According to the string-net rule given in Ref.\onlinecite{Levin2005}, $B_{p_5}$ can be obtained by merging a semion loop into plaquette $5$. Now with the link $s$ occupied by a semion string, direct calculation shows that
\be
B_{p_5} = i\prod_{k \in {p_5}} X_k \prod_{j \in {p_5}} (-)^{n_j\bar{n}_{j+1}} Z_t
\ee
where $t$ is the lower half of the link between plaquette $1$ and $5$. We can explicitly check that $B_{p_5}$ is Hermitian, $(B_{p_5})^2=I$, it commutes with all other plaquette terms already in the Hamiltonian (in the sector where all $A_v$ constraints are satisfied), but is NOT time reversal invariant. Indeed, we find
\be
\cT B_{p_5} \cT^{-1} = -B_{p_5}
\ee
To indicate the time reversal violation, we denote this term with an empty circle in Fig.\ref{doublet}.

To allow the semion string to end without violating time-reversal symmetry, we are therefore forced to remove terms containing $B_{P5}$ from the Hamiltonian. This suggests that four terms need to be removed: $B_{p_i}Z_{i5}B_{p_5}$, $i=1,2,3,4$. In fact, we can recombine the Hamiltonian terms and add some terms back (see Fig.\ref{doublet})
\be
\sum_{i=1,2,3} B_{p_i}Z_{i5}Z_{(i+1)5}B_{p_{i+1}}
\ee
These terms are time reversal symmetric, commute with all other terms in the original Hamiltonian and take eigenvalue $1$ in the ground state. Now counting the number of terms we find that we are missing one term and hence have a local two-fold degeneracy. 

Is this the local Kramer degeneracy we are looking for? Obviously this two fold degeneracy corresponds to eigenstates of $B_{p_5}$ (with eigenvalue $\pm 1$) which commutes with all other terms in the Hamiltonian but is also independent of them. Notice that $B_{p_5}$ anti-commutes with $\cT$, hence $\cT$ interchanges the two states with $B_{p_5}=\pm 1$. Therefore, to determine the $\cT^2$ value on this local degeneracy, we need to find a local operator $Q$ which maps between these two states and check its transformation under time reversal\cite{Levin2012}. That is, $Q$ needs to commute with all terms in the Hamiltonian but anti-commute with $B_{p_5}$. One possible choice is
\be
Q = B_{p_1}Z_{t}
\ee
Because
\be
(\cT Q\cT^{-1})Q = B_{p_1}Z_{t}B_{p_1}Z_{t} = -1
\ee
we see that the degenerate states form a local Kramer pair under time reversal. Or in other words, on the boundary with $\alpha=\pi$ condensate, semions transform as Kramer doublets under time reversal symmetry.

%%%%%%%%%%%%%%%%%%%%%%%%%%%%%%%%%%%%%%%%%%%%%%%%%%%%%%%%%%%%%%%%%%%%%%%%%%%%%%%

\section{Domain wall between two types of boundaries}
\label{DWall}

Although the two types of condensates, with $\alpha=0$ and $\alpha=\pi$, give rise to different boundaries with the double semion state, they do not correspond to different phases. Indeed, both $\alpha=0$ and $\alpha=\pi$ condensates are short range entangled states with time reversal symmetry. As we know that there is no nontrivial time reversal symmetry protected topological phase in 2D\cite{Chen2013a}, both condensates belong to the same phase. Therefore, the interface between these two types of condensates can be gapped out without breaking time reversal.

\begin{figure}[htbp]
\centering
\includegraphics[width=1.5in]{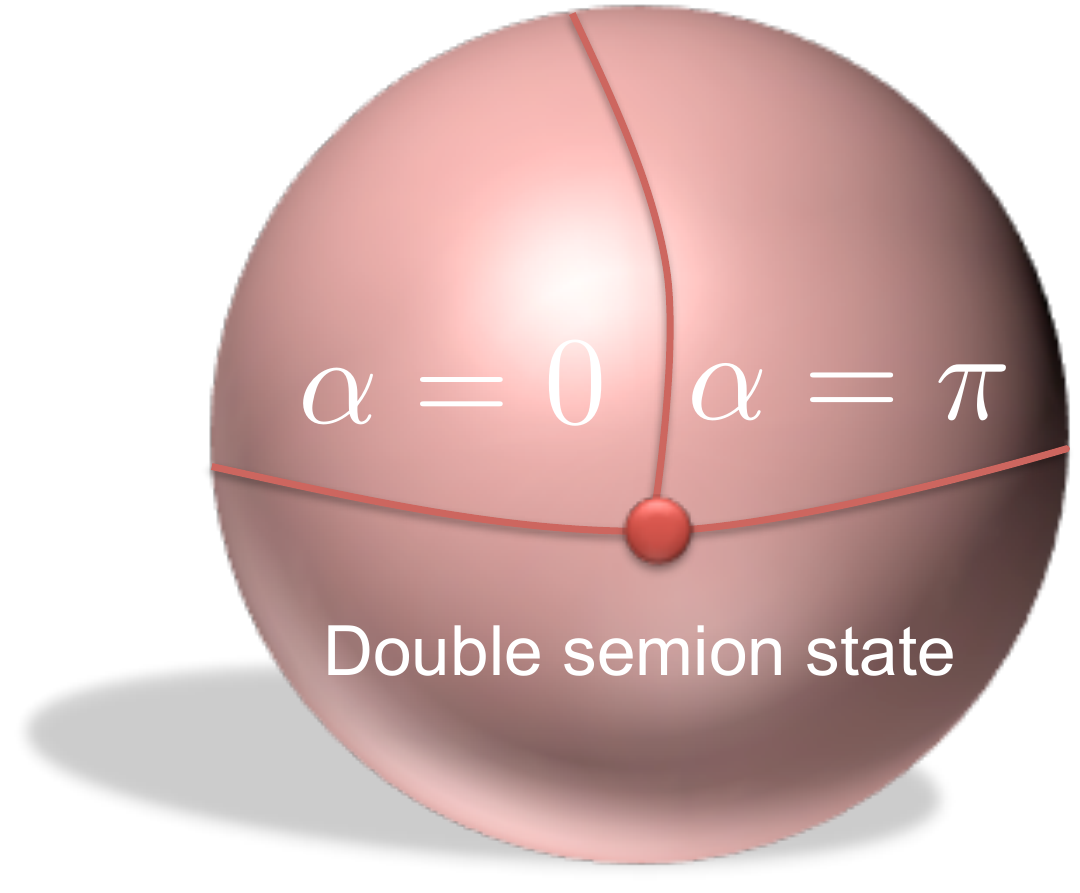}
\caption{Domain wall (red dot) between the two types of boundary between the double semion state and the $\alpha=0$ and $\alpha=\pi$ condensates respectively.}
\label{DW}
\end{figure}

Now we can ask the question of what happens on the domain wall between the two types of boundaries. Imagine a situation as shown in Fig.\ref{DW} where a 2D sphere is partitioned into three parts, occupied by the double semion state, $\alpha=0$ condensate and $\alpha=\pi$ condensate respectively. The interface between any two parts is gapped with time reversal symmetry being preserved and now we can investigate the property of the two domain walls (red dots in Fig.\ref{DW}). In this section, we are going to see whether there are degeneracies associated with the domain walls, and how they transform under time reversal symmetry.

\subsection{Field theory analysis}

From the field theory analysis in section \ref{FT}, we see that the boundary with the $\alpha=0$ condensate is in state $|A\rangle = |\phi_1-\phi_2=0\rangle$ (or equivalently $|\bar{A}\rangle = |\phi_1-\phi_2=\pi\rangle$) and the boundary with the $\alpha=\pi$ condensate is in state $|B\rangle = |\phi_1-\phi_2=\pi/2\rangle$ (or equivalently $|\bar{B}\rangle = |\phi_1-\phi_2=-\pi/2\rangle$). The difference between $|A\rangle$ and $|\bar{A}\rangle$ (or $|B\rangle$ and $|\bar{B}\rangle$) is an artificial one as they are related by the gauge transformation
\be
g: \phi_1 \rightarrow \phi_1+\pi, \phi_2 \rightarrow \phi_2
\ee
In reality, the gauge symmetry is not broken so we need to restore the symmetry and write the two boundary states as
\be
|A\rangle + |\bar{A}\rangle
\ee
and
\be
|B\rangle + |\bar{B}\rangle
\ee

When the boundary contains both parts, there are two possible configurations, 
\be
|\psi_1\rangle = |AB\rangle + |\bar{A}\bar{B}\rangle 
\ee
and
\be
|\psi_2\rangle = |A\bar{B}\rangle + |\bar{A}B\rangle 
\ee
which form a two fold degeneracy on the boundary as long as time reversal symmetry is not broken. This can be shown as follows. 
Suppose the two domain walls between type $A$ and type $B$ boundaries are at $L$ and $-L$ respectively, then the local operator near $L$ 
\be
O_1=ie^{i(\phi_1(L-\epsilon)-\phi_2(L-\epsilon))}e^{-i(\phi_1(L+\epsilon)-\phi_2(L+\epsilon))}
\ee 
($\epsilon$ is small and finite) tunnels a boson across the domain wall at $L$ and takes $\pm 1$ eigenvalues in the two states.
\be
O_1|\psi_1\rangle = |\psi_1\rangle, \ \ O_1|\psi_2\rangle = -|\psi_2\rangle
\ee 
However, $O_1$ cannot be added to the Hamiltonian because it breaks time reversal symmetry, either in the form $\cT_1$ or $\cT_2$. 
\be
\cT_1^{-1} O_1\cT_1 = -O_1, \ \ \cT_2^{-1} O_1\cT_2 = -O_1
\ee
On the other hand, the operator $O_2$ which tunnels a semion from $-L$ to $L$
\be
O_2 = e^{-i\phi_1(-L)}e^{i\phi_1(L)}
\ee
maps between $|\psi_1\rangle$ and $|\psi_2\rangle$ and anti-commutes with $O_1$. $O_2$ is a nonlocal operator and cannot be added to the Hamiltonian to split the degeneracy. Moreover, as $O_1$ and $O_2$ generate the full operator algebra of the two dimensional space spanned by $|\psi_1\rangle$ and $|\psi_2\rangle$, we see that $|\psi_1\rangle$ and $|\psi_2\rangle$ are necessarily degenerate if time reversal symmetry is preserved.

We can take either form of the time reversal action, $\cT_1$ or $\cT_2$, and we find that their action on these two states is the same as $O_2$
\be
\cT |\psi_1\rangle = |\psi_2\rangle
\ee
Therefore, the domain walls between the two types of boundaries carry a two fold degeneracy and time reversal action in this degenerate subspace is equivalent to the tunneling of semions between the two domain walls. 

More generally, if the boundary is divided into $2N$ alternating segments with $2N$ domain walls in between, similar analysis shows that there is a $2^{2N-1}$ fold degeneracy protected by time reversal symmetry. 

We are going to confirm this conclusion with exactly solvable models in the next section.

%Note that their combination 
%\be
%|\psi_1\rangle + |\psi_2\rangle = |AB\rangle + |\bar{A}\bar{B}\rangle + |A\bar{B}\rangle + |\bar{A}B\rangle
%\ee
%is not a short range correlated boundary state, for the following reason. Suppose that the two domain walls between boundaries of type $A$ and type $B$ are at $x=L$ and $x=-L$ respectively. The operator 
%\be
%O_1 = e^{i(\phi_1(L-\epsilon)-\phi_2(L-\epsilon))}e^{i(\phi_1(L+\epsilon)-\phi_2(L+\epsilon))}
%\ee 
%is a local operator for small finite $\epsilon$ and so is 
%\be
%O_2 = e^{i(\phi_1(-L-\epsilon)-\phi_2(-L-\epsilon))}e^{i(\phi_1(-L+\epsilon)-\phi_2(-L+\epsilon))}
%\ee
%Direct calculation shows that $O_1$ and $O_2$ have a finite correlation in state $|\psi_1\rangle + |\psi_2\rangle$ for arbitrary $L$
%\be
%<O_1O_2>-<O_1><O_2> = -1
%\ee
%while the same calculation gives $0$ for $|\psi_1\rangle$ and $|\psi_2\rangle$ individually. Therefore, we should only consider short range correlated states $|\psi_1\rangle$ and $|\psi_2\rangle$.

\subsection{Interface between the two types of condensates}
\label{interface}

First we need to show that the interface between the two types of condensates can be gapped. This is expected because the two condensates are both time reversal invariant short range entangled states. Because there are no nontrivial symmetry protected topological phases with time reversal symmetry in two dimensions, the two condensates are in the same phase and should be able to connect to each other in a gapped way without breaking the symmetry

\begin{figure}[htbp]
\centering
\includegraphics[width=2.8in]{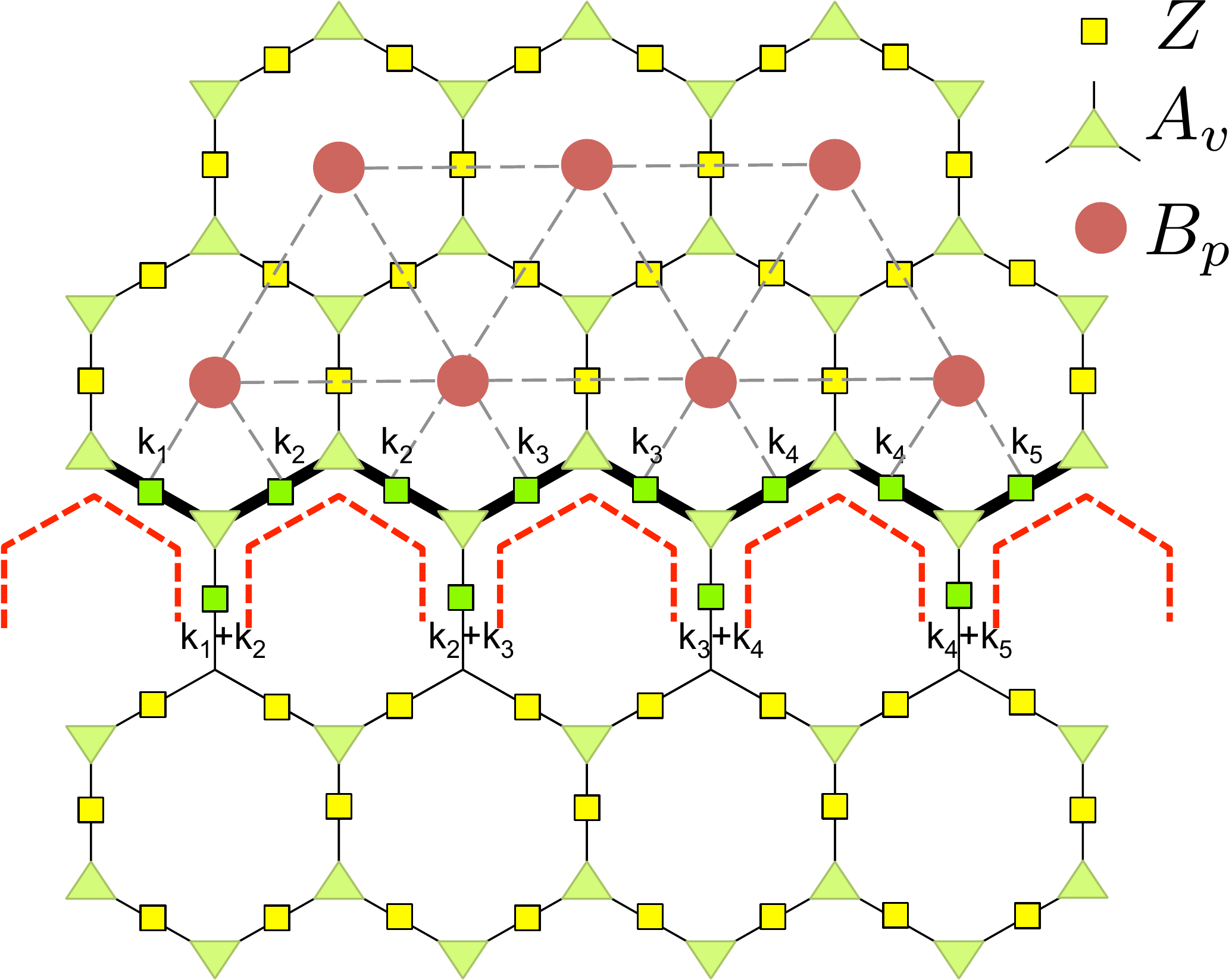}
\caption{Interface between condensate I and II.}
\label{C1C2}
\end{figure}

Consider an interface between the two condensates as shown in Fig.\ref{C1C2}

The upper half plane is in the $\alpha=\pi$ condensate with Hamiltonian terms $B_{p_a}Z_{ab}B_{p_b}$ (red dot -- yellow square -- red dot) and the lower half plane is in the $\alpha=0$ condensate with Hamiltonian terms $Z_{k}$ (yellow square). 
The green squares label the degrees of freedom on the interface and form the interface Hilbert space. 

Let's be more explicit about this.  First, notice that the $\alpha=\pi$ condensate can be mapped to the $\alpha=0$ condensate by unitary transformation 
\be
U=\prod_{p\in C_{\pi}} \frac{(I+iB_p)}{1+i}
\ee
where the product is over all placates in the condensate. We can see this explicitly from
\be
\begin{array}{l}
\frac{(I+iB_{p_a})}{1+i}\frac{(I+iB_{p_b})}{1+i}Z_{ab}\frac{(I-iB_{p_a})}{1-i}\frac{(I-iB_{p_b})}{1-i} \\
= B_{p_a}Z_{ab}B_{p_b}
\end{array}
\ee
Therefore, after applying $U$, all links inside the two condensates (yellow squares) are in the state $|n_i=0\rangle$ and $U\prod|0\rangle$ is the ground state of these regions. 
 
Next, we enforce the closed loop constraints
\be
A_v=\prod_{i\in v} Z_i
\ee
along the interface (thick black line) as indicated by the green triangles.

Now we are left with one 2 dimensional degree of freedom per plaquette, labelled by $k_s=0,1$ as shown in Fig.\ref{C1C2}. $|k_s=0\rangle$ and $|k_s=1\rangle$ are eigenvalue $1$ and $-1$ eigenstates of $Z_s$ respectively. Each state in the interface Hilbert space can be then be written as
\be
|\psi_{\{k_s\}}\rangle = U \prod_{\triangledown} |k_s\rangle |k_{s+1}\rangle |k_s+k_{s+1}\rangle \prod_C |0\rangle
\label{psi_b}
\ee
where the first product is over all downward pointing vertices on the interface and the second product is over all the links inside the two condensates (the ones with the yellow squares). Note that the links on the interface are not all independent due to the $A_v$ constraints. There is one free degree of freedom per plaquette on the interface. As given in Eq.\ref{psi_b}, each state $|\psi_{\{k_s\}}\rangle$ is the eigenvalue $e^{i\pi k_s}$ eigenstate of $iZ_sB_{p_{s}}$, where $s$ labels links along the interface and $p_s$ labels the plaquette on the $\alpha=\pi$ condensate side of link $s$. Under time reversal, which acts as complex conjugation, $iZ_sB_{p_{s}}$ is mapped to $-iZ_sB_{p_{s}}$. Therefore, each $|\psi_{\{k_s\}}\rangle$ is not time reversal invariant.

To find a time reversal symmetric interface, we must add some terms to the boundary that mix the $|\psi_{k_s} \rangle$ states, respect $\cT$, and gap the interface out. While it may be hard to directly find such an operator, we can apply the unitary transformation $U=\prod_{p \in C_{\pi}} \frac{(I+iB_p)}{1+i}$ and map the interface Hilbert space to that spanned by
\be
|\psi^0_{\{k_s\}}\rangle = \prod_{\triangledown} |k_s\rangle |k_{s+1}\rangle |k_s+k_{s+1}\rangle \prod_C |0\rangle
\ee
The transformed interface Hilbert space now takes a simple direct product form of local degrees of freedom labeled by $k_s$ and allows simpler analysis of possible Hamiltonian terms. Even though the unitary $U$ is not local, it preserves the spectrum and hence a gapped edge in this basis is also gapped in the original basis.

There is one complication though: The action of time reversal is also transformed under $U$.  Before we can write down time reversal invariant Hamiltonians, we need to find the correct time reversal transformation in this new basis. 
\be
\tilde{\cT} = U^{\dagger}\cT U = \prod_{p\in C_{\pi}} \left(\frac{1+iB_p}{1+i}\right)^2 = \prod_{p\in C_{\pi}} B_p
\ee
Therefore, the effective time reversal action $\tilde{\cT}$ on the interface Hilbert space is complex conjugation in the $|\psi^0_{\{k_s\}}\rangle$ basis and
\be
\prod X_s \prod (-)^{n_s(1-n_{s+1})}
\ee
Although it looks complicated and non-onsite, we know that it should have a short range entangled ground state. Indeed we find that under local unitaries
\be
V = \prod_s (i)^{n_s(1-n_{s+1})} \ \ \ .
\ee
The effective time reversal action is mapped to
\be
V\tilde{\mathcal{T}}V^{\dagger} = \prod X_s K
\ee
which has a gapped symmetric Hamiltonian $\sum_s X_s$. Note that in Fig.\ref{C1C2} $X_s$ acts on four links, the two links labeled by $k_s$ and also the ones labeled by $k_{s-1}+k_s$ and $k_s+k_{s+1}$. This is illustrated in Fig.\ref{C1C2} with red dotted lines.

Therefore, the transformed interface Hilbert space spanned by $|\psi^0_{\{k_s\}}\rangle$ can be gapped in a time reversal invariant way by effective Hamiltonians
\be
\sum_s V^{\dagger}X_sV
\ee
In the original interface Hilbert space spanned by $|\psi_{\{k_s\}}\rangle$, the Hamiltonian reads
\be
\sum_s UV^{\dagger}X_sVU^{\dagger}
\ee
In appendix \ref{real}, we explicitly confirm that the term $UV^{\dagger}X_sVU^{\dagger}$ is indeed real.

\subsection{Tri-junction between $\alpha=0$, $\alpha=\pi$ condensates and double semion}

Now we have found a gapped interface between the $\alpha=0$ and the $\alpha=\pi$ condensates, we can study the domain wall between the two types of boundaries as the tri-junction between the two condensates and the topological state (double semion).

\begin{figure}[htbp]
\centering
\includegraphics[width=3.0in]{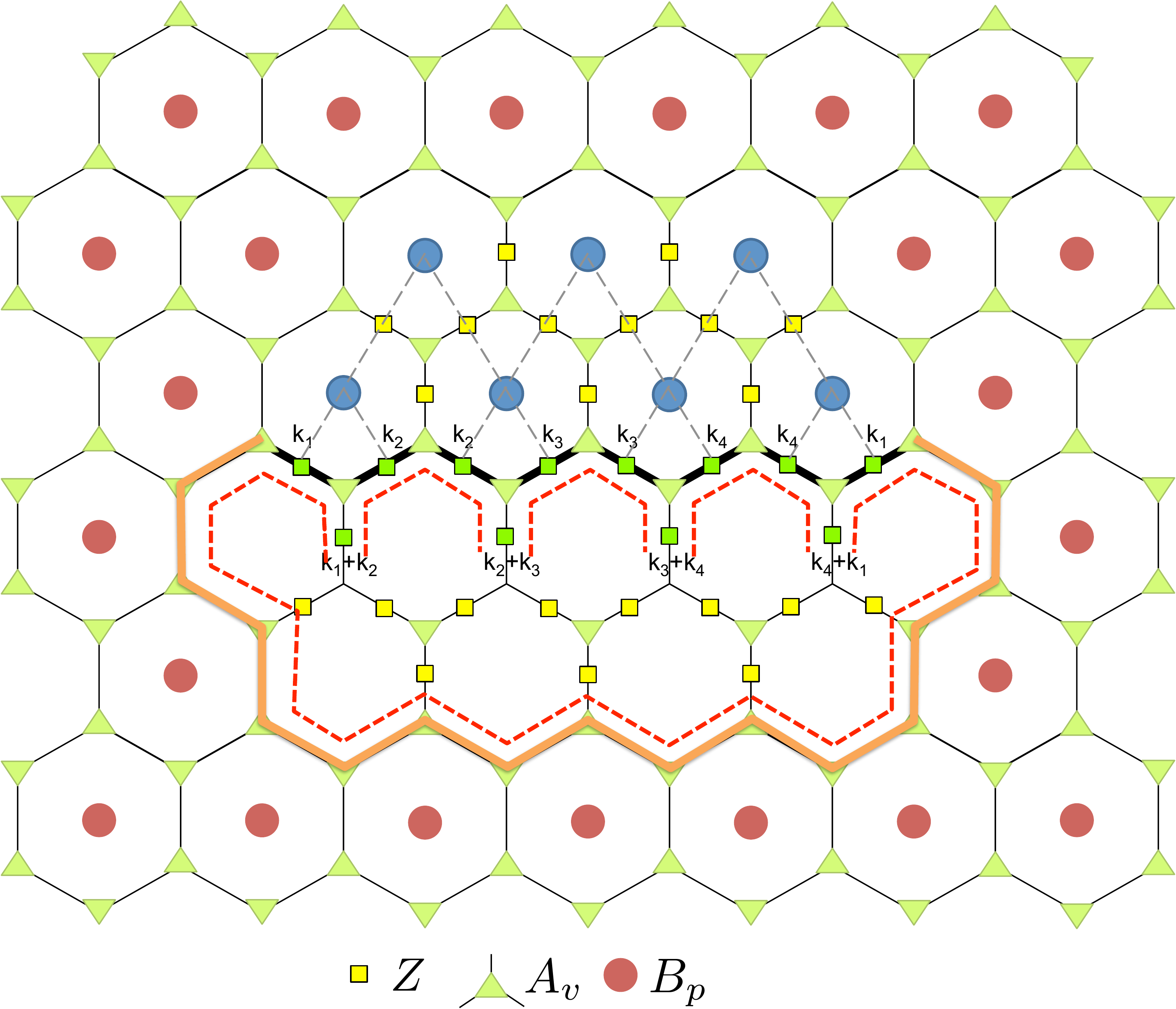}
\caption{Tri-junction between $\alpha=0$, $\alpha=\pi$ condensates and the double semion state.}
\label{C1C2DS}
\end{figure}

As shown in Fig. \ref{C1C2DS}, topological region is on the outside and the two condensates are in the middle. Following previous discussion, Hamiltonian in the topological region is given by (red dots and green triangles)
\be
H_T = \sum_v A_v + \sum_p B_p
\ee
Hamiltonian in the $\alpha=0$ condensate is given by (yellow squares)
\be
H_{C_0} = -\sum_k Z_k -\sum_v A_v
\ee
and that in the $\alpha=\pi$ condensate is given by (blue dot -- yellow square -- blue dot)
\be
H_{C\pi} =-\sum_{<ab>} B_{p_a}Z_{ab}B_{p_b} - \sum_v A_v
\ee
Moreover, we enforce the closed loop constraints $A_v$ along the interface between the two condensates (thick black line).

Now we can see what the low energy Hilbert space is composed of. Apply again the unitary transformation $U=\prod_{p\in C_{\pi}} \frac{1+iB_p}{1+i}$. Note that $U$ commutes with all the terms in $H_T$ and the $A_v$ constraints on the interface, maps the $B_{p_a}Z_{ab}B_{p_b}$ terms in $H_{C\pi}$ to $Z_{ab}$ and leaves $H_{C0}$ untouched. Therefore, states in the low energy Hilbert space can be written as
\be
|\phi_{\{k_i\}}\rangle = \prod_{p \in C_{II}} \frac{I+iB_p}{1+i} |\phi^0_{\{k_i\}}\rangle
\ee
where
\be
|\phi^0_{\{k_i\}}\rangle = |\phi_{DS}\rangle \prod_{\triangledown} |k_i\rangle |k_{i+1}\rangle |k_i+k_{i+1}\rangle \prod_C |0\rangle 
\ee
The first product is over all downward pointing triangles on the interface between the two condensates and the second product is over all links in the two condensates. Note that the two $k_1$'s near the two tri-junctions are the same, due to the closed loop constraints over the whole state.
$|\phi_{DS}\rangle$ is the wave function in the topological region (including all the links around red dots as shown in Fig.\ref{C1C2DS}). The exact form of $|\phi_{DS}\rangle$ depends on whether $k_1=0$ or $k_1=1$. The two can be mapped into each other by running a semion string operator from one tri-junction to another along, for example, the orange line shown in Fig.\ref{C1C2DS}. 

Following a similar line of reasoning as discussed in the previous section, we find that the interface can be gapped with time reversal invariant terms
\be
UV^{\dagger}X_sVU^{\dagger}
\ee
where $U$ is again a product over plaquettes in the $\alpha=\pi$ condensate $U=\prod_{p \in C_{\pi}} \frac{1+iB_p}{1+i}$ and $V=\prod_s (i)^{n_s(1-n_{s+1})}$.

There is one major difference though  from the situation discussed in the previous section. $X_1$ is now a nonlocal operator. Not only does $X_1$ change $k_1$ to $1-k_1$, it also changes the form of $|\phi_{DS}\rangle$ by running a semion string from one tri-junction to another. Therefore, if we require time reversal symmetry and locality, we are left with a two fold degeneracy with $k_1=0$ or $1$ respectively. Time reversal symmetry maps $k_1$ to $1-k_1$ which can be equivalently accomplished by $X_1$ in this degenerate Hilbert space. Therefore, time reversal symmetry acts as semion tunneling in the low energy Hilbert space of tri-junctions.

%%%%%%%%%%%%%%%%%%%%%%%%%%%%%%%%%%%%%%%%%%%%%%%%%%%%%%%%%%%%%%%%%%%%%%%%%%%%%%%

\section{Relation to $Z_2\times Z_2^T$ symmetry protected topological phase}
\label{SPT}

The discussions in the previous sections tell us that there is one symmetry enriched topological phase for double semion topological order with time reversal symmetry. However, there can be two different gapped boundaries with time reversal symmetry. The double semion topological order is a twisted $Z_2$ gauge theory and by un-gauging the $Z_2$ symmetry we can get $Z_2\times Z_2^T$ symmetry protected topological phases. So how is our conclusion about double semion SET consistent with what we know about $Z_2\times Z_2^T$  SPT?

From the classification of SPT we know that there are four phases with $Z_2\times Z_2^T$ symmetry\cite{Chen2013a}. In particular, there are two root phases: phase $(1,0)$ which has nontrivial $Z_2$ SPT order and trivial action of time reversal and phase $(0,1)$ with trivial $Z_2$ SPT order and projective action of time reversal on $Z_2$ twist defects\cite{Chen2014a}. Phase $(1,1)$ is their combination with both nontrivial $Z_2$ SPT order and projective $Z_2$ twist defects under time reversal.

After gauging the $Z_2$ symmetry, we get SET phases. Phase $(0,0)$ and $(0,1)$ lead to the usual non-twisted $Z_2$ gauge theory (toric code order) and the resulting states differ in the way $Z_2$ fluxes transform under time reversal (as singlet or doublet). Therefore, after gauging, phase $(0,0)$ and $(0,1)$ results in two different SET phases. 

On the other hand,  phase $(1,0)$ and $(1,1)$ gauge into the twisted $Z_2$ gauge theory (double semion order). As there is only one time reversal SET with double semion order, the two SPT phases have to merge into one upon gauging. To understand how this happens we notice that phase $(1,0)$ and $(1,1)$ can be mapped into each other by relabeling the group element. In particular, in the symmetry group $Z_2\times Z_2^T = \{I,g,\cT,g\cT\}$, there are two anti-unitary operators $\cT$ and $g\cT$. If we exchange the two, the group is still a $Z_2\times Z_2^T$ group and phase $(1,0)$ is mapped into phase $(1,1)$ and vice verse. Therefore, upon gauging the $Z_2$ symmetry, phase $(1,0)$ and $(1,1)$ become the same SET.

More explicitly, the edge state of the $Z_2 \times Z_2^T$ SPTs can be described in general with
\be
L_e = \frac{1}{2\pi} \partial_x\phi_1 \partial_t \phi_2
\ee
Different phases correspond to different ways symmetry act on $\phi_1$ and $\phi_2$. In both phase $(1,0)$ and $(1,1)$, $Z_2$ acts as
\be
g: \phi_1 \rightarrow \phi_1 + \pi, \phi_2 \rightarrow \phi_2 + \pi
\ee
Time reversal acts respectively as
\be
\begin{array}{l}
\cT_1: \phi_1 \rightarrow \phi_1, \phi_2 \rightarrow -\phi_2 \\
\cT_2: \phi_1 \rightarrow \phi_1 + \pi, \phi_2 \rightarrow -\phi_2
\end{array}
\ee
$e^{i(\phi_1 + \phi_2)/2}$ (or $e^{i(\phi_1-\phi_2)/2}$) creates a $Z_2$ twist defect on the edge and transforms as $\cT_1^2=1$ and $\cT_2^2=-1$. 

Combining $g$ with $\cT_1$ we get
\be
g\cT_1: \phi_1 \rightarrow \phi_1+\pi, \phi_2 \rightarrow -\phi_2-\pi
\ee
Redefine $\tilde{\phi_2} = \phi_2 +\pi/2$, we find
\be
g\cT_1: \phi_1 \rightarrow \phi_1+\pi, \tilde{\phi_2} \rightarrow -\tilde{\phi_2}
\ee
which is exactly the same as the action of $\cT_2$. Hence after gauging $\cT_1$ and $\cT_2$ are equivalent, and both correspond to the same SET phase. However a memory of these two distinct SPT phases is retained at the boundary of our SET, where depending on boundary conditions the semion can transform with either $\cT^2 = 1$ or $-1$.

%%%%%%%%%%%%%%%%%%%%%%%%%%%%%%%%%%%%%%%%%%%%%%%%%%%%%%%%%%%%%%%%%%%%%%%%%%%%%%%

\section{Conclusion and Discussion}
\label{Conclusion}

To summarize, in this paper we have learned the following things about the double semion topological order:
\begin{enumerate}
\item There are two different ways to condense boson and gap out the boundary while preserving time reversal symmetry in the double semion state. One corresponds to a Bose condensate with a coherent phase factor $1$ and the semion excitations on the boundary transform as $\cT^2=1$. The other corresponds to a Bose condensate with a coherent phase factor of $-1$ and the semion excitations on the boundary transform as $\cT^2=-1$.
\item A pair of domain walls between the two types of boundaries carry a two fold degeneracy. Time reversal symmetry acts in this two dimensional space by tunneling a semion from one domain wall to another.
\item There is only one SET phase with time reversal symmetry and double semion topological order. The different transformation properties of the semions under time reversal symmetry is a pure boundary effect. In the bulk, semions are mapped to anti-semions and it is not meaningful to talk about the time reversal representation carried by semion itself. 
\item Different SPT phases can become the same SET phase after partly gauging the unitary symmetry of the system. Examples of this kind have been pointed out in Ref.\onlinecite{Cheng2014}. 
\end{enumerate}

These results can be generalized to other SET or SPT phases. In appendix \ref{SPTSET}, we present another example of SPTs coalescing upon gauging with unitary symmetry. We leave the study of more general cases to the future. 

We want to comment briefly on the relation between this double semion example and some previous studies of gapped boundaries of topological states. 

A simple yet very interesting case was the boundary state of toric code topological order. Even in the absence of symmetry, there are two types of boundaries corresponding to the two types of (self) bosons in the toric code\cite{Bravyi1998}. Each domain wall between the two types of boundaries carry a Majorana mode, giving rise to a $2^N$ fold degeneracy for $N+1$ pairs of domain walls (with fixed fermion parity). 

Similarly, degeneracies arise with domain walls in our double semion example. With $N$ pairs of domain walls, there is a $2^{2N-1}$ fold degeneracy. Of course, this degeneracy requires the protection of time reversal symmetry and can be completely removed by adding time reversal symmetry breaking local terms. If time reversal symmetry is preserved, the degenerate states can only be mapped to each other through non-local operators which tunnel semionic excitations from one domain wall to another. 

The existence of such domain wall degeneracy has also been noticed in (fractional) topological insulators where domain walls between ferromagnetic gapped edges and superconducting gapped edges carry Majorana (parafermion) zero modes\cite{Fu2009,Cheng2012,Clarke2013,LIndner2012,Vaezi2013}. Such models are different from the double semion example studied here in that symmetry is broken in order to gap out the edge.

Finally,  Wang and Levin studied different ways to gap out the edge of a `strong pairing insulator'\cite{Wang2013b}, either with an interface with a topological insulator or with an interface with a trivial insulator. Symmetry is preserved in these two kinds of edges, but the interface between the two condensates -- the topological insulator and the trivial insulator -- is always gapless when symmetry is preserved.

%%%%%%%%%%%%%%%%%%%%%%%%%%%%%%%%%%%%%%%%%%%%%%%%%%%%%%%%%%%%%%%%%%%%%%%%%%%%%%%

\acknowledgments

We would like to thank Lukasz Fidkowski, Zhenghan Wang, Meng Cheng, and T. Senthil for discussion. XC is supported by the Miller Institute for Basic Research in Science at UC Berkeley, the Caltech Institute for Quantum Information and Matter and the Walter Burke Institute for Theoretical Physics. AV is supported by the Templeton Foundation.

%\bibliographystyle{apsrev_nurl}
%\bibliography{TRB_DS_bib}

\appendix

\section{Explicit calculation of $\cT$-invariance of the gapping terms between the two condensates}
\label{real}

In this section, we are going to explicitly verify that the $UV^{\dagger}X_sVU^{\dagger}$ term given in section \ref{interface} is indeed real.

\begin{figure}[htbp]
\centering
\includegraphics[width=1.5in]{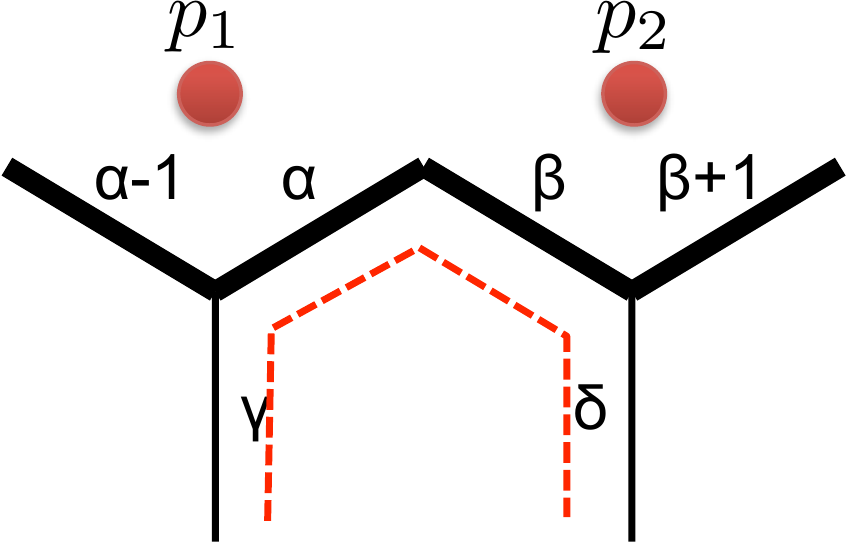}
\caption{Each $X_s$ term flips four links $\alpha$, $\beta$, $\gamma$ and $\delta$.}
\label{Xs}
\end{figure}

First, let's write the $X_s$ term in full as $\sigma^x_{\gamma}\sigma^x_{\alpha}\sigma^x_{\beta}\sigma^x_{\delta}$, as shown in Fig.\ref{Xs}
Then
\be
\begin{array}{ll}
&V^{\dagger}X_sV  \\
= & S_{\alpha-1}^{\dagger}S_{\beta}^{\dagger}\sigma^x_{\gamma}\sigma^x_{\alpha}\sigma^x_{\beta}\sigma^x_{\delta}S_{\alpha-1}S_{\beta} \\
= & \sigma^x_{\gamma}\sigma^x_{\alpha}\sigma^x_{\beta}\sigma^x_{\delta} V_{\phi}
\end{array}
\ee
where 
\be
S_{a}=i^{n_{a}(1-n_{a+1})}
\ee
and
\be
V_{\phi}=i^{n_{\alpha-1}+1-n_{\beta+1}}(-)^{n_{\alpha-1}n_{\alpha}+(1-n_{\beta})(1-n_{\beta+1})}
\ee

Complex conjugating $UV^{\dagger}X_sVU^{\dagger}$ then gives
\be
\begin{array}{ll}
& (UV^{\dagger}X_sVU^{\dagger})^* \\
= & \frac{1-iB_{p_1}}{1-i}\frac{1-iB_{p_2}}{1-i}\sigma^x_{\gamma}\sigma^x_{\alpha}\sigma^x_{\beta}\sigma^x_{\delta} V^*_{\phi}\frac{1+iB_{p_1}}{1+i}\frac{1+iB_{p_2}}{1+i} \\
= & UB_{p_1}B_{p_2}\sigma^x_{\gamma}\sigma^x_{\alpha}\sigma^x_{\beta}\sigma^x_{\delta} V^*_{\phi}B_{p_1}B_{p_2}U^{\dagger}
\end{array}
\ee
where $p_1$ and $p_2$ are the plaquettes above link $\alpha-1$, $\alpha$ and $\beta$, $\beta+1$ respectively, as shown in Fig.\ref{Xs}.

Therefore, $UV^{\dagger}X_sVU^{\dagger}$ is real if
\be
B_{p_1}B_{p_2}\sigma^x_{\gamma}\sigma^x_{\alpha}\sigma^x_{\beta}\sigma^x_{\delta} V^*_{\phi}B_{p_1}B_{p_2} = \sigma^x_{\gamma}\sigma^x_{\alpha}\sigma^x_{\beta}\sigma^x_{\delta} V_{\phi}
\ee
This can be explicitly checked as
\be
\begin{array}{lll}
B_{p_1}B_{p_2}\prod \sigma^x &= & \prod \sigma^x (-)^{n_{\alpha-1}+n_{\beta+1}+1} \\
(-)^{n_{\alpha-1}+n_{\beta+1}+1}V_{\phi}^*&=&V_{\phi}
\end{array}
\ee
and $B_{p_1}B_{p_2}$ commutes with $V_{\phi}$. Therefore, 
\be
(UV^{\dagger}X_sVU^{\dagger})^* = UV^{\dagger}X_sVU^{\dagger}
\ee

\section{Example of SPTs coalescing upon gauging with unitary symmetry}
\label{SPTSET}

Consider SPTs with $Z^{(1)}_2 \times Z^{(2)}_2$ (unitary) group in 2d. The classification is $Z^3_2$. The root phases have the edge structure,
\beq L = \frac{i}{2\pi} \d_x \phi \d_\tau \theta\eeq
and transformation properties:
\bea  \mathrm{Phase} \,\,(1,0,0): && Z^{(1)}_2:\quad \phi \to \phi+\pi, \quad \theta \to \theta + \pi, \nn\\
&& Z^{(2)}_2: \quad trivial\nn\\
\nn\\
\mathrm{Phase} \,\,(0,1,0): && Z^{(1)}_2:\quad trivial\nn\\&& Z^{(2)}_2:\quad \phi \to \phi+\pi, \quad \theta \to \theta + \pi
\nn\\
\nn\\
\mathrm{Phase} \,\,(0,0,1): && Z^{(1)}_2:\quad \phi \to \phi + \pi, \quad \theta \to \theta \nn\\&& Z^{(2)}_2:\quad \phi \to \phi, \quad \theta \to \theta + \pi\nn
\eea

Combining phases $(0,0,1)$ and $(0,1,0)$ we obtain 
\bea \mathrm{Phase} \,\,(0,1,1): && Z^{(1)}_2:\quad \phi \to \phi + \pi, \quad \theta \to \theta \nn\\&& Z^{(2)}_2:\quad \phi \to \phi + \pi, \quad \theta \to \theta + \pi
\label{eq:011}\eea

Indeed, denote the edge modes of $(0,0,1)$ as $\phi_1, \theta_1$ and edge-modes of $(0,1,0)$ as $\phi_2, \theta_2$. Then the combined edge is described by,
\beq L = \frac{i}{2\pi} \d_x \phi_1 \d_\tau \theta_1 + \frac{i}{2\pi} \d_x \phi_2 \d_\tau \theta_2 \eeq
Let's define $\phi'_1 = \phi_1, \,\theta'_1 = \theta_1 - \theta_2,\, \phi'_2 = \phi_1 + \phi_2, \,\theta'_2 = \theta_2$.  The action has the same form in the primed variables as in the unprimed. 
The transformation properties of the primed variables are,
\bea 
Z^{(1)}_2:\quad &&\phi'_1 \to \phi'_1 + \pi, \quad \theta'_1 \to \theta'_1 \nn\\
&&\phi'_2 \to \phi'_2 + \pi, \quad \theta'_2 \to \theta'_2 \nn\\
Z^{(2)}_2:\quad &&\phi'_1 \to \phi'_1 , \quad \theta'_1 \to \theta'_1 \nn\\
&&\phi'_2 \to \phi'_2 + \pi, \quad \theta'_2 \to \theta'_2 +\pi \nn\\
\eea
Adding a term $-\lambda \cos(\theta'_1)$ we gap out the $\phi'_1$, $\theta'_1$ modes. The transformation properties of $\phi'_2, \theta'_2$ are then exactly the same as in Eq.~(\ref{eq:011}).]

Let us denote the generator of  $Z^{(1)}_2$ as $g_1$, the generator of $Z^{(2)}_2$ as $g_2$ and $g_3 = g_1 g_2$. Then, in phase $(0,1,1)$ under $g_3$ we have
\beq g_3: \quad \phi \to \phi, \quad \theta \to \theta + \pi\eeq
So the phase $(0,1,1)$ is like the phase $(0,0,1)$ but with the actions of $g_2$ and $g_3$ interchanged. Now imagine gauging the $Z^{(1)}_2$ group in the two phases $(0,0,1)$ and $(0,1,1)$. 
The action of $g_1$ in both cases is identical and the twist defect is given by $e^{i \tilde{\theta}} = e^{i \theta/2}$. In terms of $\tilde{\theta}$ the edge theory is,
\beq  L = \frac{2 i}{2\pi} \d_x \phi \d_\tau \tilde{\theta}\eeq
i.e. after gauging we get a toric code topological order. Now,  $Z^{(2)}_2$ remains a global symmetry in the resulting SET,
\bea (0,0,1)\,\, \mathrm{with}\,\,Z^{(1)}_2\,\mathrm{gauged}; \ && Z^{(2)}_2: \phi \to \phi, \tilde{\theta} \to \tilde{\theta} + \pi/2\nn\\
(0,1,1)\,\, \mathrm{with}\,Z^{(1)}_2\,\,\mathrm{gauged}; \ && Z^{(2)}_2: \phi \to \phi + \pi, \tilde{\theta} \to \tilde{\theta} + \pi/2\nn\\
\eea
Clearly, as SETs the two phases are the same since all local degrees of freedom transform in the same way (the transformation properties differ by a pure ``gauge" transformation $\phi \to \phi + \pi$). 

We can also think about gauging the remaining $Z^{(2)}_2$ symmetry. In the case of $(0,0,1)$ the twist defect of $g_2$ is $e^{i \tilde{\phi}} = e^{i \phi/2}$. Thus, we get an overall $Z_4$ topological order,
\beq  L = \frac{4 i}{2\pi} \d_x \tilde{\phi} \d_\tau \tilde{\theta}\eeq
where $e^{i \tilde{\phi}}$ is the twist defect of $Z^{(2)}_2$ and $e^{i \tilde{\theta}}$ is the twist defect of $Z^{(1)}_2$. 

Now, in the case of $(0,1,1)$ the twist defect of $g_2$ is $e^{i \phi/2} e^{i \theta/2} = e^{i \tilde{\phi}} e^{i \tilde{\theta}}$. If we are interested in the ``overall" topological order,
we can still use the $e^{i \tilde{\phi}}$, $e^{i \tilde{\theta}}$ basis, obtaining a $Z_4$ topological order. However, the way that the twist defect of $g_2$ is ``embedded" within this topological order
is different. In $(0,0,1)$ it is just $e^{i \tilde{\phi}}$ (which is a boson), whereas in $(0,1,1)$ it is $e^{i \tilde{\phi} }e^{i \tilde{\theta}}$ (which is a semion). If we are treating the two phases as SET
phases (with $Z^{(1)}_2$ ``fully gauged"), then the difference of an extra factor of $e^{i \tilde{\theta}}$ is irrelevant, since the twist defect of the global symmetry $Z^{(2)}_2$ can always trap an extra anyon (vison) of the SET, $e^{i \tilde{\theta}}$.
Thus, we cannot distinguish the two SET phases.  However, if we are working with SPT phases and putting the twist defects in ``by hand", then we know whether the twist defect of $g_1$ is present or not, so can distinguish the two SPT phases.

\end{document}